# On the prediction of critical heat flux using a physics-informed machine learning-aided framework

Xingang Zhao[a], Koroush Shirvan[a,*], Robert K. Salko[b], Fengdi Guo[c]

[a] *Department of Nuclear Science and Engineering, Massachusetts Institute of Technology, Cambridge, MA 02139, USA*

[b] *Reactor and Nuclear Systems Division, Oak Ridge National Laboratory, Oak Ridge, TN 37831, USA*

[c] *Department of Civil and Environmental Engineering, Massachusetts Institute of Technology, Cambridge, MA 02139, USA*

[*] Corresponding author. *E-mail address*: kshirvan@mit.edu (Koroush Shirvan)

**Abstract**

The critical heat flux (CHF) corresponding to the departure from nucleate boiling (DNB) crisis is essential to the design and safety of a two-phase flow boiling system. Despite the abundance of predictive tools available to the thermal engineering community, the path for an accurate, robust CHF model remains elusive due to lack of consensus on the DNB triggering mechanism. This work aims to apply a physics-informed, machine learning (ML)-aided hybrid framework to achieve superior predictive capabilities. Such a hybrid approach takes advantage of existing understanding in the field of interest (i.e., domain knowledge) and uses ML to capture undiscovered information from the mismatch between the actual and domain knowledge-predicted target. A detailed case study is carried out with an extensive DNB-specific CHF database to demonstrate (1) the improved performance of the hybrid approach as compared to traditional domain knowledge-based models, and (2) the hybrid model's superior generalization capabilities over standalone ML methods across a wide range of flow conditions. The hybrid framework could also readily extend its applicability domain and complexity on the fly, showing an elevated level of flexibility and robustness. Based on the case study conclusions, the window-type extrapolation mapping methodology is further proposed to better inform high-cost experimental work.

**Nomenclature**

| | |
|---|---|
| $D$ | diameter |
| $D_e$ | channel equivalent (or hydraulic) diameter |
| $D_h$ | channel heated diameter |
| $G$ | mass flux |
| $G_{cutoff}$ | cutoff mass flux |
| $h_{fg}$ | latent heat of vaporization |
| $L_B$ | vapor blanket length |
| $L_h$ | heated length |
| $n$ | number of observations in the dataset |
| $P$ | pressure |
| $q''_{chf,exp}$ | experimental CHF |
| $U_B$ | vapor blanket velocity |
| $\boldsymbol{x}$ | input feature vector |
| $x_e$ | equilibrium (or thermodynamic) quality |
| $y$ | actual (measured) output |
| $\hat{y}_h$ | predicted output with hybrid framework |
| $\hat{y}_p$ | predicted output with prior model |

*Greek letters*

| | |
|---|---|
| $\delta$ | liquid sublayer thickness |
| $\varepsilon$ | residual |
| $\hat{\varepsilon}_m$ | predicted residual with ML |
| $\rho_f$ | liquid density at saturation |

*Abbreviations*

| | |
|---|---|
| ANN | artificial neural network |
| API | application programming interface |
| CHF | critical heat flux |
| DK | domain knowledge |
| DNB | departure from nucleate boiling |
| EPRI | Electric Power Research Institute |
| LUT | look-up table |
| MAE | mean absolute error |
| ML | machine learning |
| MSE | mean squared error |
| NN | (feed-forward) neural network |
| PWR | pressurized water reactor |
| ReLU | rectified linear unit |
| RF | random forest |
| rRMSE | relative root-mean-square error |

# 1. Introduction

The reliability and economic competitiveness of a thermal system hinge upon its safety and regulatory measures. During the stage of system design and analysis, researchers and engineers typically leverage extensive experimental efforts and investigate evolutionary models that represent the state-of-the-art understanding in their fields of specialization—also known as *domain knowledge* (DK)—to predict various safety limits. However, engineering problems with sophisticated physical phenomena may present extreme challenges to establishing explicit mathematical expressions or building credible input/output causality. Therefore, such safety limits are often determined by overly conservative measures, as they are accompanied with significant margins to accommodate modeling errors and other potential uncertainties. One such engineering problem is the *departure from nucleate boiling* (DNB) crisis in a two-phase flow boiling system, widely encountered in high-power microprocessor cooling, in the refrigeration industry, in some medical technology fields, and more typically in nuclear power plants [1]. The DNB crisis is characterized by a sharp deterioration of the heat transfer coefficient at the heater/coolant interface, as vapor permanently blankets the heated surface, preventing access of supply liquid (which is subcooled or slightly saturated) and potentially leading to a set of cascading component failures. The corresponding heat flux, or *critical heat flux* (CHF), is a regulatory limit for commercial pressurized water reactors (PWRs) worldwide [2].

Unfortunately, the path for an accurate, robust prediction of CHF has been elusive due to lack of general agreement on the mechanism that triggers DNB [3]. Despite the disagreement, the thermal engineering community has proposed a large number of DK-based predictive tools, from correlations and look-up tables (LUTs) to physics-based mechanistic models [1,4]. On one hand, the data-driven best-fit correlations and LUTs are easy to implement but highly empirical. They may result in relatively good agreement with specific experimental datasets but often fail to extend beyond their ranges of validity in terms of geometry and operating conditions [3]. Commonly used empirical CHF models include the Zuber correlation [5], the Biasi correlation [6], the Groeneveld 2006 LUT [7], the W-3 correlation [8], and the Electric Power Research Institute (EPRI) correlation [9]. On the other hand, the physics-based models rely on assumptions from reasonable yet limited understanding of the underlying physics and are supplemented with mostly empirical constitutive relations to close the conservation equations. Since the mid-1960s, numerous mechanistic DNB models in flow boiling have been developed and are generally grouped into six categories [1] based on their main respective triggering mechanisms. Among these categories, the liquid sublayer dryout has received considerable attention due to experimental evidence obtained with internally heated round tubes [1,10].

With recent advances in computational capabilities and optimization techniques, methods based on *machine learning* (ML) provide an alternative approach to existing data-driven DK-based tools. Such an approach can be particularly useful in engineering fields where the physical phenomena are complex. Within this category, an artificial neural network (ANN) is one of many promising choices, as it has been shown to serve as a universal approximator of any nonlinear relations [11,12]. Depending on its expected outcomes, different types of ANNs have been applied to a variety of disciplines, including those related to applied thermal engineering: from predicting solar radiation [13] and wind speed [14] to forecast of pressure drop in heat exchangers [15]. Applications in nuclear engineering have also increased but often face one key obstacle: lack of experimental or high-fidelity numerical data. In regard to the prediction of CHF, most nonproprietary measurements were conducted with steam-water mixtures in round tubes. None of the few publications in this field [4,16–18] attempted to distinguish DNB from dryout, a thermal crisis triggered at a much higher quality and lower heat flux [19]. The term CHF is sometimes confusingly used for both boiling crises, though. No recent publications employed cross-validation to assess their network architectures, and none evaluated the sensitivity of different hyperparameters (i.e., parameters with values that are set before learning starts) or discussed regularization, an ML technique that prevents overfitting and reduces test error. Another popular supervised ML branch for regression, the tree-based ensemble learning, excels in established high efficiency and robustness [20], although its engineering applications have been scarce to date.

While standalone ML-based tools require minimal a priori knowledge and almost no explicit mathematical modeling, they can be prone to undesired, unphysical solutions due to their purely data-driven nature and "black-box" feature. Regardless of how advanced any ML method has become, prior knowledge in the field (i.e., DK) is still deemed important and useful by many researchers [21–25]. Accordingly, the idea of combining ML and DK arose through the concept of a hybrid/integrated "gray-box" framework, first in chemical process industries [21–24] and later for applications in electrical engineering [25] and aerospace engineering [26]. The "gray-box" framework can be further classified into two main types: the *series* approach, in which ML is applied to estimate intermediate variables that are key closures in a physics-driven DK-based model; and the *parallel* approach, also known as the physics-informed, ML-aided framework, which uses ML to compensate the bias of DK-based models and to capture the undiscovered information from the mismatch between the actual and DK-predicted target. Regarding the prediction of DNB-type CHF, the series "gray-box" approach could be applied to a liquid sublayer dryout mechanistic model, where ML would help forecast such closure terms as bubble velocity, drag coefficient, and bubble departure diameter. However, at this stage, a much more extensive database is required for proper ML training on any of these intermediate variables, so the series approach is not explored here.

This work extends prior work on CHF by offering a comprehensive assessment of the physics-informed, ML-aided framework (or *hybrid framework* for short). Section 2 introduces the framework and describes the dataset used in the case study. Section 3 briefs preliminary work on standalone ML in the first place and then details performance of the hybrid framework, including choice of DK model and extrapolation capabilities. The window-type extrapolation mapping methodology is further proposed to help inform future experiments. Finally, Section 4 summarizes the main conclusions and presents future work.

## 2. Methodology and dataset
### 2.1. Physics-informed, ML-aided framework

In this hybrid parallel framework, a conventional DK-based model is selected as the fixed-structure prior model, either data- or physics-driven. While DK serves to lay the groundwork and provide a baseline solution, ML is used to learn from the residual between actual and DK-predicted output (not from the final output as it is the case of a standalone ML-based model). Figure 1 presents its workflow during training and validation/testing. Input information is available to both the prior model and the ML method. For CHF applications, the input feature vector $x$ usually comprises (at least) six (6) variables: pressure ($P$), mass flux ($G$), local equilibrium (or thermodynamic) quality ($x_e$), channel equivalent (or hydraulic) diameter ($D_e$), channel heated diameter ($D_h$), and heated length ($L_h$). The prior model's predicted output ($\hat{y}_p$) is essentially a nonlinear function of input features, $f(x)$, in the form of an empirical correlation, table, or system of first-principle equations. At the training stage (Fig. 1a), the residual ($\varepsilon$) is obtained by subtracting $\hat{y}_p$ from the actual/measured output ($y$), and ML is then trained on the residual. The ML-predicted residual ($\hat{\varepsilon}_m$) is compared with $\varepsilon$ through a loss (or cost) function. The objective of the training process is to optimize (minimize) the loss function, usually presented as the mean squared error (MSE) or the mean absolute error (MAE). The performance of the final predicted output ($\hat{y}_h$)—sum of $\hat{y}_p$ and $\hat{\varepsilon}_m$—is evaluated against the experimental CHF ($y = q''_{chf,exp}$) by means of the relative root-mean-square error (rRMSE), defined as:

$$rRMSE = \sqrt{\frac{1}{n}\sum_{i=1}^{n}(\frac{y^{(i)}-\hat{y}_h^{(i)}}{y^{(i)}})^2}, \tag{1}$$

where $n$ is the number of observations in the dataset. Similarly, at the validation/test stage (Fig. 1b), the prior model and ML are combined to determine the predicted output.

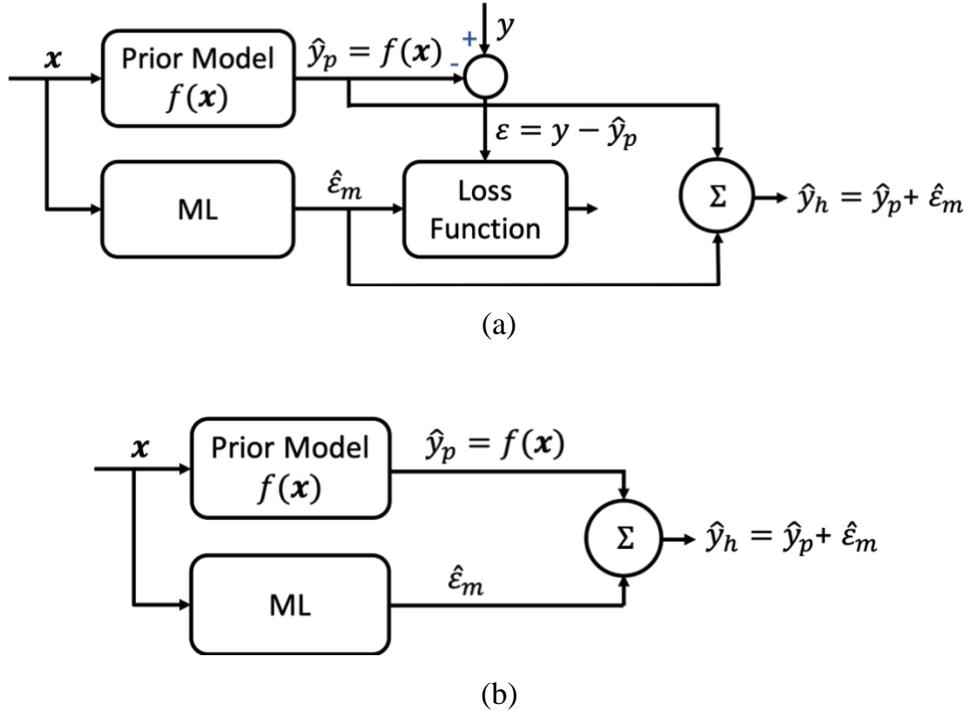

**Fig. 1.** Simplified structure for hybrid framework during (a) training and (b) validation/testing.

### 2.1.1 Prior/DK model

Two prior models are selected for this work: the Groeneveld 2006 LUT [7], one of the most prevalent and accurate data-driven tools for predicting CHF in current nuclear thermal-hydraulics community; and the Liu model [27], one of the most recent and successful in the series of physics-driven tools based on the relatively well-accepted liquid sublayer dryout mechanism.

*Look-up table (LUT)*

The Groeneveld CHF LUT [7] is a normalized data bank for a vertical 8 mm water-cooled round tube. The 2006 version is based on a database containing over 30,000 data points, covering the full range of conditions of practical interest: $0.1 \leq P \leq 21$ MPa, $0 < G \leq 8,000$ kg/m²-s, and $-0.50 < x_e \leq 0.90$. Table values are adjusted by multiplicative correction factors to account for different diameter/geometry/heat flux distribution conditions [2,7,28]. In noncircular channels (e.g., annular, rectangular, rod bundle subchannel), the heated diameter ($D_h$) rather than the hydraulic diameter ($D_e$) is recommended for making the diameter correction, as the former feature better describes vapor formation and development in subcooled and low-quality flow [29]. Note that the LUT is applicable to both DNB and dryout scenarios. The table method is simple to use and has a very low computational cost.

*Mechanistic Liu model*

The Liu model [27] is based on the liquid sublayer dryout theory, which assumes that the onset of DNB is caused by the complete evaporation of a thin superheated liquid layer underneath a vapor blanket flowing over the heated wall. The vapor blanket is formed as a consequence of coalescing small bubbles rising along the near-wall region. Therefore, using heat balance, a simplified governing equation can be written as:

$$CHF = \frac{\rho_f \delta h_{fg}}{L_B} U_B, \qquad (2)$$

where $\rho_f$ is the liquid density at saturation and $h_{fg}$ is the latent heat of vaporization; $\delta$, $U_B$ and $L_B$ are respectively the liquid sublayer thickness, vapor blanket velocity, and vapor blanket length. The key for the liquid sublayer dryout mechanism turns to the determination of these three parameters (i.e., $\delta$, $U_B$ and $L_B$), along with related intermediate variables. A variety of physics-based models are available in the literature [3,10,27,30,31], showing promising insights as well as limitations. As one of the most recent in the series, the Liu model focused on analyzing instabilities at both the interface of the liquid sublayer/vapor blanket and that of the vapor blanket/bulk region. Although the underlying mechanism was insensitive to channel geometry, considerations were only given to round tubes for closure relations in almost all the existing mechanistic models (including Liu). Liu et al. [27] reported that their model compared favorably with its previous counterparts for its validation matrix (2,482 tube data), although the statistics were clearly outperformed by those of the data-driven LUT. The model was found deficient for predictions at low subcooling conditions.

### 2.1.2 ML method

Among the various types of ANNs that exist, the *feed-forward neural network* (or *multilayer perceptron*, or simply *NN* in this article) has become the most popular in engineering applications for its superior performance when dealing with strongly nonlinear and complex relationships. Another ML method being used in this case study is the *tree-based random forest* (*RF*), a fast and reliable tool that requires minimal hyperparameter tuning and feature engineering.

*Feed-forward neural network (NN)*

The feed-forward neural network is essentially a collection of multilayer (at least three layers: input, hidden, output), fully-connected units (see Fig. 2 for a sample architecture) capable of nonlinear mapping via activation functions between two layers. Weights and biases are randomly initiated from a uniform distribution and iteratively updated during training by the backward propagation algorithm using gradient descent. To ensure generalization (i.e., to prevent overfitting), various regularization techniques such as dropout [32] or early-stopping [33] can be applied to help reduce the test error.

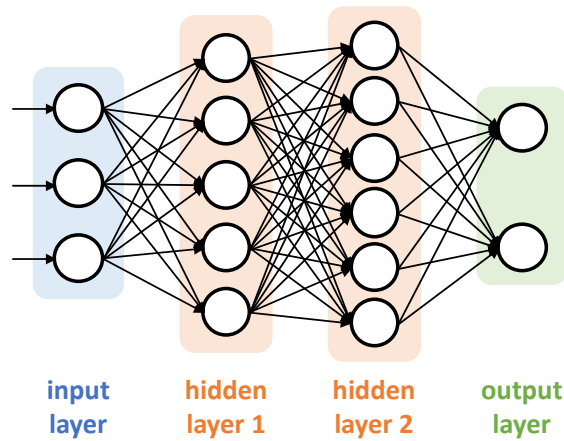

**Fig. 2.** Typical feed-forward NN: example of a 3/5/6/2 architecture.

*Random forest (RF)*

The random forest is a fast, flexible, tree-based ensemble learning method that produces robust results without much tuning of hyperparameters [20]. By randomly selecting observations and features with the bootstrap aggregation technique (also known as *bagging*, a model averaging approach that is designed to improve accuracy, prevent overfitting, and reduce variance), multiple decision trees are aggregated, and their predictions are then averaged, as depicted in Fig. 3.

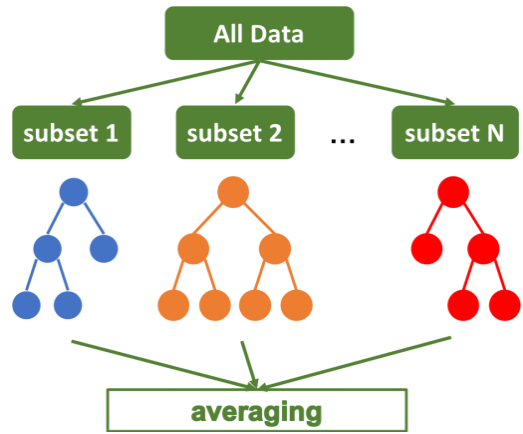

**Fig. 3.** Simplified RF structure.

### 2.2. Dataset description

Relevant publications [34–43] which included tabulated nonproprietary experimental CHF data have been reviewed, and a total of 1,865 test cases were collected. The dataset covers a wide range of flow conditions while limiting the focus to DNB-specific characteristics (i.e., local/exit[a] equilibrium quality $\leq 0.2$). The heaters are either round tubes, annuli, or one-side heated plates. As mentioned in Section 2.1, the input features collected from raw data consist of pressure, mass flux, local/exit equilibrium quality, channel equivalent diameter, channel heated diameter, and heated length. The target/output is CHF, and the axial power profile is uniform. The experimental ranges of the dataset are summarized in Table 1. It should be noted that no geometry indicator is included in the baseline input feature vector (i.e., the ML algorithm does not know whether an observation is linked with a tube, an annulus, or a plate). The sensitivity of such an indicator will be briefly evaluated in Appendix A. The experimental uncertainties were estimated[a] to be in the range of 5–15% [42,43].

---

[a] With a uniform axial heat flux distribution, DNB first occurs at the channel outlet.

**Table 1**

CHF dataset: value ranges of baseline input features.

| Author | Geometry | Feature #1 *pressure* [MPa] | Feature #2 *mass flux* [kg/m$^2$-s] | Feature #3 *equilibrium quality* [-] | Feature #4 *equivalent diameter* [mm] | Feature #5 *heated diameter* [mm] | Feature #6 *heated length* [mm] | # of data |
|---|---|---|---|---|---|---|---|---|
| *Inasaka* [34] |  | 0.31 to 0.91 | 4,300 to 6,700 | -0.15 to -0.04 | 3.0 | 3.0 | 100 | 7 |
| *Peskov* [35] |  | 12 to 20 | 750 to 5,361 | -0.23 to 0.13 | 10.0 | 10.0 | 400 to 1,650 | 17 |
| *Thompson* [36] | TUBE | 0.1 to 20.7 | 542 to 7,975 | -0.45 to 0.20 | 1.0 to 37.5 | 1.0 to 37.5 | 25 to 3,048 | 1,202 |
| *Weatherhead* [37] |  | 13.8 | 332 to 2,712 | -0.49 to 0.19 | 7.7 to 11.1 | 7.7 to 11.1 | 457 | 162 |
| *Williams* [38] |  | 5.5 to 15.2 | 670 to 4,684 | -0.03 to 0.17 | 9.5 | 9.5 | 1,836 | 51 |
| *Beus* [39] |  | 5.5 to 15.5 | 671 to 3,721 | -0.31 to 0.20 | 5.6 | 15.2 | 2,134 | 77 |
| *Janssen* [40] | ANNULUS | 4.1 to 9.7 | 381 to 5,913 | -0.13 to 0.20 | 4.6 to 22.2 | 11.3 to 96.3 | 737 to 2,743 | 282 |
| *Mortimore* [41] |  | 8.3 to 13.8 | 677 to 3,637 | -0.13 to 0.20 | 5.0 | 13.3 | 2,134 | 19 |
| *Kossolapov* [42] | PLATE | 0.1 | 350 to 2,078 | -0.14 to -0.02 | 15.0 | 120.0 | 10 | 12 |
| *Richenderfer* [43] |  | 0.1 to 1.0 | 1,000 to 2,000 | -0.04 to -0.01 | 15.0 | 120.0 | 10 | 36 |

As a key step in the machine learning pipeline, feature engineering is the process of using a priori expertise to transform raw data into a set of features with properties that can be effectively handled by ML algorithms [44]. While the RF approach generally performs well with raw inputs, NN requires feature scaling. For this work, standardization (calculated using the arithmetic mean and standard deviation of the given data, which normalizes inputs to have zero mean and unity variance) is applied to the input vectors of pressure, mass flux, equivalent diameter, heated diameter, and heated length.

### 2.3. ML training and validation

Both NN and RF methods are trained and validated using the Keras application programming interface (API) with TensorFlow[b] backend and the scikit-learn library in Python 3.6. The 10-fold cross-validation technique is used to tune the algorithms and evaluate how well they generalize on unseen data. The dataset is randomly shuffled prior to cross-validation. The shuffled dataset is then evenly divided into ten (10) subsamples/folds: nine (9) for training and one (1) for validation. This process is repeated ten (10) times, with each of the ten (10) subsamples used exactly once as the validation fold. Given the relatively small size of the dataset, a separate test fold is not generated. This evaluation technique will be compared with the more conventional train–test split in Section 3.3 for out-of-sample predictions.

---

[a] The measurement uncertainties were not reported in most aforementioned references.
[b] TensorFlow is an open-source symbolic tensor manipulation framework developed by Google.

## 3 Results and discussion

### 3.1 Preliminary study: best-estimate standalone ML vs. LUT

Previous work by Zhao [45] discussed the effects of different algorithms and hyperparameters on the performance of a standalone NN, as well as regularization, and reached the following conclusions that apply to this case study:

- *weight optimization* and *learning rate*: Adam optimizer[a] marginally outperformed the classic stochastic gradient descent procedure, optimal learning rate = 0.001;
- *activation function* (in hidden layers): the rectified linear unit (ReLU)[b] compared favorably to sigmoid[c];
- *network architecture*: optimal configuration = 6/50/50/50/1 (the input layer has 6 units, each of the 3 hidden layers has 50 units, and the output layer has 1 unit); more hidden layers or units did not further reduce validation error;
- *number of epochs*: convergence was reached after 600–800 epochs[d];
- *regularization*: the validation error never increased with the number of epochs, and applying dropout (an efficient regularization technique that refers to randomly dropping out units during training [32]) did not help further reduce the validation error. Therefore, overfitting should not be a concern in this work.

Similarly, hyperparameter sensitivity in a standalone RF was explored, and only minor changes in its performance were observed (i.e., RF requires minimal tuning). In summary, the best-estimate standalone ML-based CHF models comprised the following configuration:

---

[a] The Adam optimization algorithm [52] combines the advantages of multiple extensions of the classic stochastic gradient descent procedure. Derived from the adaptive moment estimation, it is popular in the field of deep learning and requires minimal tuning.
[b] $ReLU(x) = max(0, x)$ is a universal approximator, as any function can be approximated with combinations of ReLU. Its downside—the zero-gradient issue with no weight updates—was found insignificant for this application. Its variant, LeakyReLU, did not further improve the NN performance.
[c] Sigmoid is another popular NN activation function which is expressed as: $sigmoid(x) = 1/(1 + e^{-x})$.
[d] Each epoch represents one forward pass and one backward pass of all the training examples.

- *NN*: 6/50/50/50/1 architecture, Adam optimizer (learning rate = 0.001), ReLU activation, no regularization;
- *RF*: 100 trees/estimators, 50%-70% features in each individual tree, no regularization.

The comparison against the well-used LUT is detailed in Zhao et al. [46] and summarized in Appendix A. The standalone ML-based methods are found to compare favorably with the conventional DK tools in terms of flexibility and ease of modeling (i.e., minimal prior knowledge is required). A key advantage is their online extensibility of the applicability domain. However, their purely data-driven nature and "black-box" characteristics may lead to largely scattered, physically undesired solutions (illustrations from this work can be found in Section 3.2). Elaborating the root cause of such scatter is challenging: it may be due to the presence of outliers in the training and/or validation dataset; or maybe the training and validation data do not follow the same statistical distribution, a key prerequisite to any meaningful validation/test solutions in ML. Leveraging prior knowledge will help reduce such scatter, since DK is capable of providing credible baseline solutions and ML is then used to assist DK and learn from the prediction mismatch.

### 3.2 Hybrid approach vs. standalone models

The hybrid (i.e., physics-informed ML-aided) framework described in Section 2.1 has been implemented for this case study. The prior model can be either the data-driven LUT or the physics-based Liu model; the ML method can be either NN or RF. To evaluate the effect of prior models (LUT and Liu) on the performance of the hybrid approach, only tube results are compared (since the Liu model was developed for tube applications only). Hyperparameter tuning has revealed that within the hybrid framework, the ML structure can be simplified: in the NN, a single hidden layer of 30 units (i.e., 6/30/1 architecture) is sufficient to achieve the maximum performance; in the RF, only 50 trees (instead of 100) are needed. No other changes are made from the configuration presented in Section 3.1.

Figure 4 compares the performance of the standalone models (ML: NN and RF; DK: LUT in Fig. 4a and Liu in Fig. 4b) and that of the hybrid models (NN+LUT and RF+LUT in Fig. 4a; NN+Liu and RF+Liu in Fig. 4b). All ML (standalone and hybrid) results presented here are from the 10-fold cross-validation. The hybrid approach clearly outperforms the other models: the rRMSE values are much smaller, and the corresponding cumulative data fraction curves rise much faster than those with standalone ML or DK models. Within the hybrid framework, since the prior model is fixed throughout the training and validation process, the ML-aided component can guarantee (for interpolation or generalization) that the final prediction is at least as good as that obtained with the prior model stand-alone [25], regardless of the ML method's complexity. The presence of a prior model, whether data- or physics-driven, lays the groundwork for the hybrid "gray-box". Such a framework is capable of leveraging widely admitted laws of physics and well-established empirical relations in the prior model to synergize with ML. One may argue that with an ideally configured structure, a standalone ML model (especially an ANN) could theoretically emulate similar performance. However, such efforts would be deemed cumbersome and would introduce an extra level of complexity.

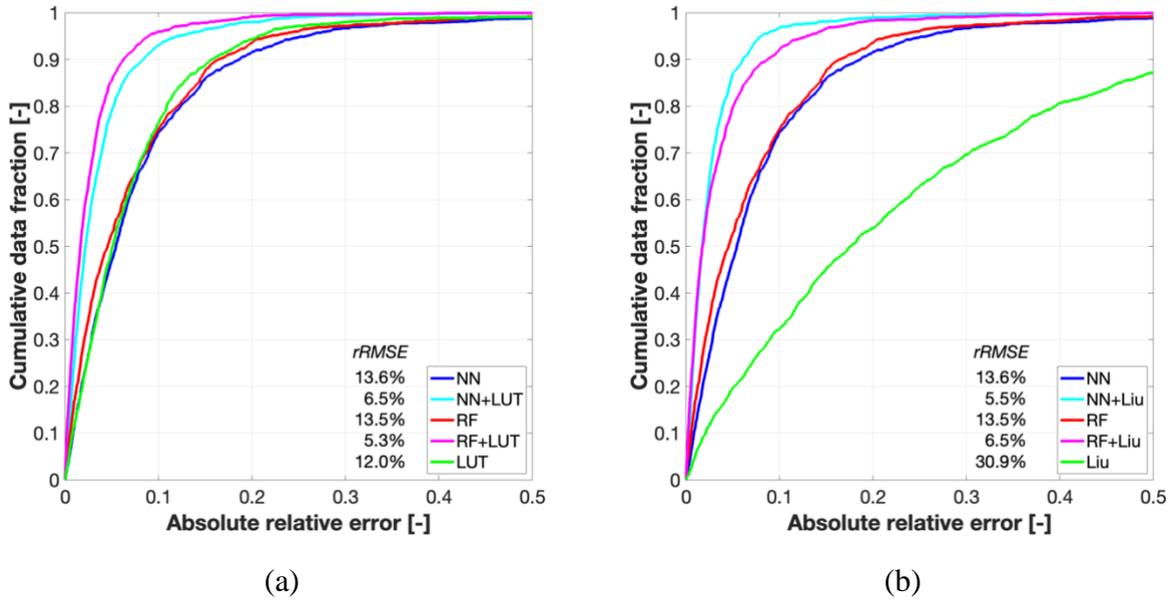

**Fig. 4.** Cumulative data fraction of absolute relative error with standalone and hybrid models on tube data: (a) LUT as DK model, (b) Liu as DK model.

Another observation from Fig. 4 is that while the standalone Liu model performs significantly worse than LUT, their hybrid version results (combined with either NN or RF) are closely in line. This implies that the ML-aided component is likely to play a larger role when dealing with more scattered and biased residuals, or when the prior model is less accurate. A reasonable interpretation is that randomness may prevail over regular undiscovered trends in the residuals if the prior model is very accurate [22], making it difficult for ML to learn the "true" information.

As shown in Fig. 5, the hybrid model (RF+LUT[a]) also corrects largely scattered and biased parametric trends (vs. pressure, mass flux, exit equilibrium quality, tube diameter, and tube length-to-diameter ratio) with standalone DK or ML models over the ranges of practical interest. Such observations have further confirmed the enhanced generalization capabilities of the hybrid approach.

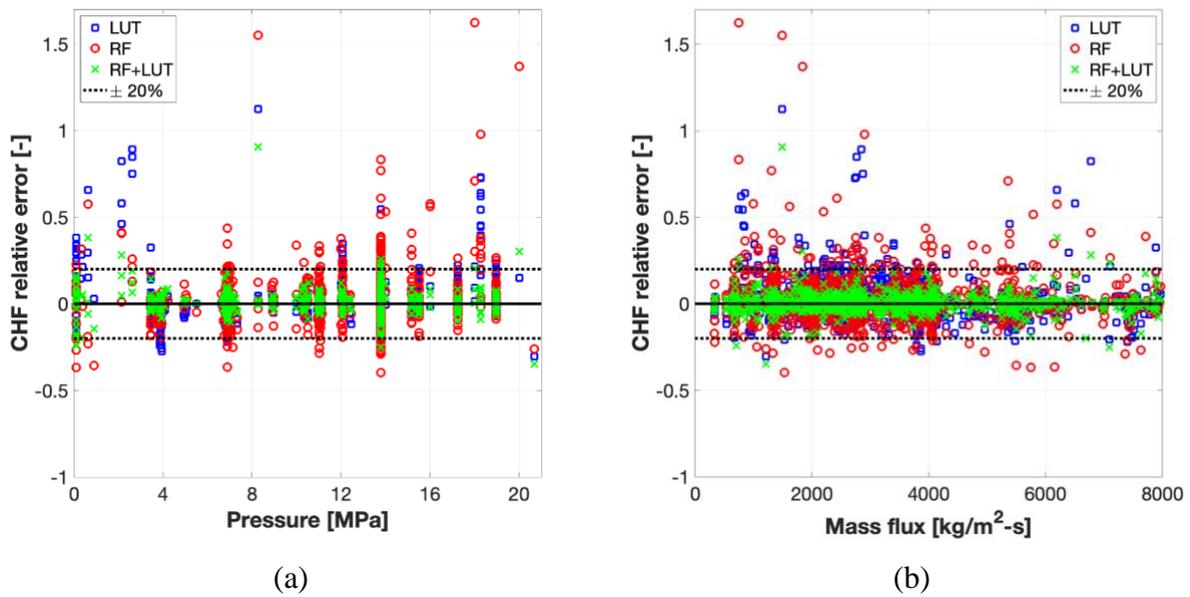

(a)                      (b)

---

[a] One can draw the same conclusion with other combinations of ML and prior model (RF+Liu, NN+LUT, NN+Liu).

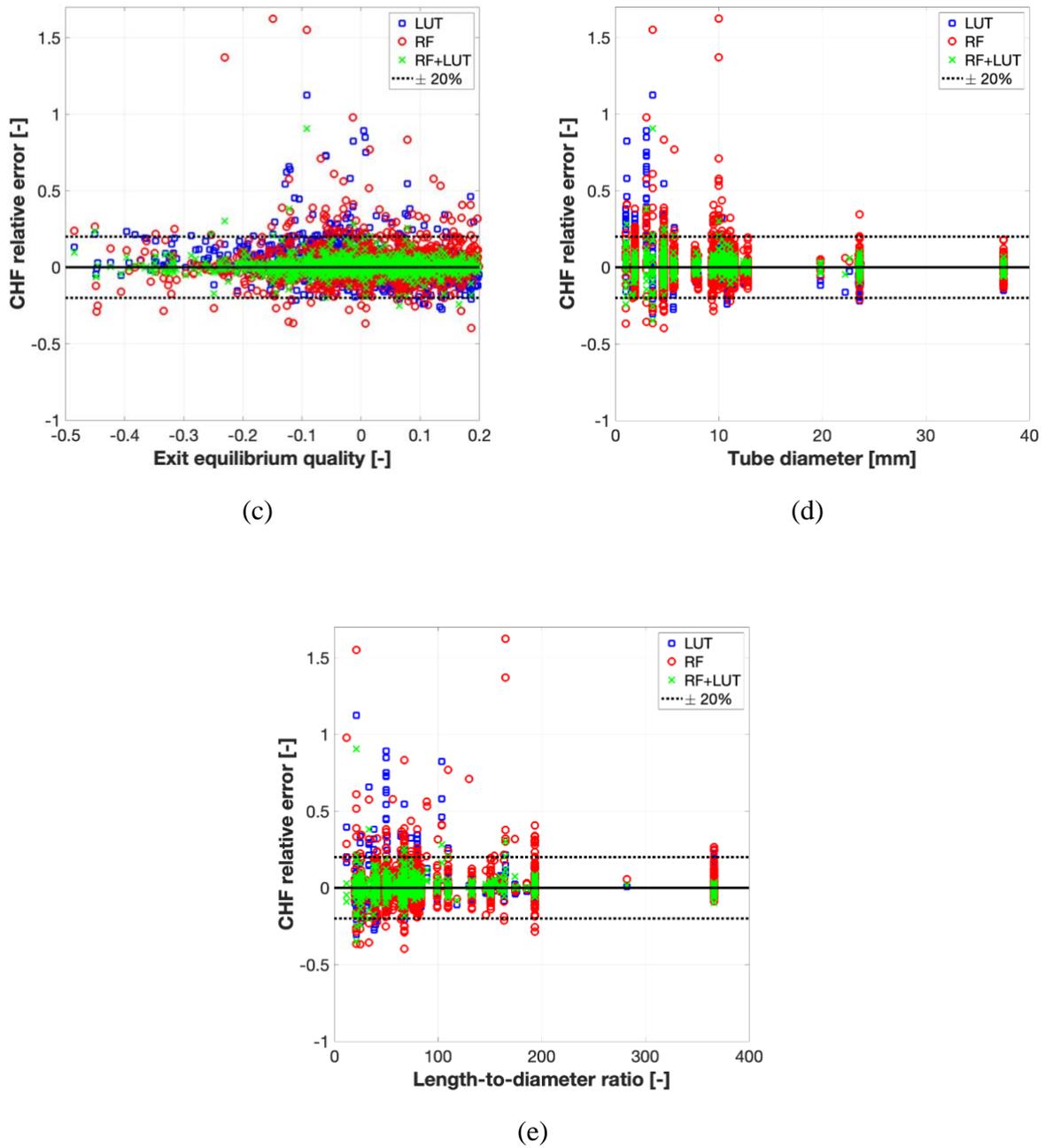

(c)

(d)

(e)

**Fig. 5.** Tube CHF relative error[a] with standalone (LUT; RF) and hybrid (RF+LUT) models vs. (a) pressure, (b) mass flux, (c) exit equilibrium quality, (d) diameter, and (e) length-to-diameter ratio.

---

[a] CHF relative error = (predicted CHF − experimental CHF) / experimental CHF.

### 3.3 Sensitivity study: ML evaluation techniques

To evaluate the ML model's performance and detect potential overfitting, both cross-validation and train–test split techniques are recommended. The latter, more straightforward train–test split was used in most prior work on CHF problems [4,16–18]. Hence, the hybrid model (RF+LUT) performance with the train–test split technique is compared against that with cross-validation in Table 2, where only minor differences are observed. While both techniques are applicable to this case study, cross-validation generates more random splits and reduces error due to bias by creating more subsets. One should also take into consideration the computational price that increases with the number of subsamples. Given the data size of this application, computational cost is not a primary concern: for a single evaluation (training plus validation) of the entire dataset on a personal computer (macOS with a 3.5 GHz Intel Core i7 processor and 16GB RAM), RF runs the fastest (< 1s), while NN (1–2 min) is slightly slower than the LUT.[a]

**Table 2**

Performance of hybrid (RF+LUT) model on tube data: sensitivity of evaluation techniques.

| Evaluation technique (all *pre-shuffled*) | Validation/test rRMSE | Data within ±10% error | Data within ±20% error |
|---|---|---|---|
| 10-fold cross validation (*baseline*) | 0.05 | 95% | 99% |
| 5-fold cross validation | 0.05 | 94% | 99% |
| 90% train + 10% test | 0.04 | 96% | 100% |
| 80% train + 20% test | 0.05 | 95% | 99% |
| 67% train + 33% test | 0.06 | 94% | 99% |

### 3.4 Extrapolation capabilities

As discussed in Section 1, a main downside of data-driven DK models is their poor extrapolation performance, as they are often incapable of making reliable predictions on data that fall beyond their validity range. For its data-driven nature, a standalone ML-based model would also be exposed to such deficiency [47].[b] Table 3 shows an example of mass flux extrapolation within the scope of this case study. The tube dataset is divided into two groups: a training group for data with a mass flux below the cutoff $G_{cutoff}$, and a test group for data with a mass flux above $G_{cutoff}$. The goal is to evaluate the performance of different ML (standalone and hybrid) methods at relatively high mass flux given that they are trained with relatively low mass flux data only. This approach is adopted since experiments at high flow velocity conditions can be costly and technically challenging in real-world engineering. Both standalone LUT and Liu serve as *reference solutions* as their validity ranges cover both training and test data in this example, that is to say that their extrapolation capabilities are not evaluated in this work.

---

[a] Note that in the hybrid framework, the predetermined prior model-related computational cost is irrelevant to the decision on evaluation technique, since the residual space is kept unchanged.

[b] In the study performed by He and Lee [47] on CHF predictions using (standalone) ML, higher pressure CHF extrapolation with low pressure data would be conditionally possible if "a few" high pressure data points were mingled in the training stage.

**Table 3**

Performance of different models on tube data: mass flux extrapolation.

| Cutoff mass flux $G_{cutoff}$ = [kg/m²-s] | | 5,000 | 4,000 | 3,000 | 2,500 | 2,000 | 1,500 | 1,000 | 800 |
|---|---|---|---|---|---|---|---|---|---|
| # of training data (for ML) | | 1,242 | 1,031 | 794 | 602 | 434 | 297 | 118 | 74 |
| # of test data (for ML) | | 197 | 408 | 645 | 837 | 1,005 | 1,142 | 1,321 | 1,365 |
| Test rRMSE | RF | 0.21 | 0.18 | 0.25 | 0.27 | 0.27 | 0.28 | 0.34 | 0.38 |
| | LUT (*reference*) | 0.12 | 0.10 | 0.10 | 0.12 | 0.12 | 0.11 | 0.12 | 0.12 |
| | RF+LUT | 0.10 | 0.07 | 0.07 | 0.09 | 0.08 | 0.11 | 0.11 | 0.12 |
| | Liu (*reference*) | 0.27 | 0.32 | 0.36 | 0.35 | 0.34 | 0.33 | 0.32 | 0.32 |
| | RF+Liu | 0.13 | 0.11 | 0.12 | 0.13 | 0.17 | 0.17 | 0.25 | 0.28 |

As listed in Table 3, the RF extrapolates poorly stand-alone but consistently improves when combined with a prior model. Unlike interpolation (see Fig. 4), the choice of DK model used in the hybrid framework becomes more important for extrapolation, especially when the training dataset is relatively small: LUT significantly outperforms Liu, and the same conclusion holds when combined with RF. Note that in this example, the extrapolation terminology only refers to the ML-aided component of the hybrid approach (the physics-informed component is not extrapolated here), i.e., in respect of the residuals. Finally, from Table 3, one can notice that the hybrid models predict at least as accurately as their standalone DK counterparts, although their extrapolation capabilities are not directly compared (again, standalone DK models do not extrapolate here).

### 3.5 Window-type extrapolation mapping

In light of the promising performance of the hybrid framework, one may leverage existing experimental data to inform whether or not new measurements are needed for a targeted thermal-hydraulic condition. Such initiatives are economically attractive for applications like DNB since measuring CHF can be expensive and time consuming [48]. Generally, decisions are made based on the performance metrics (e.g., rRMSE) of *in-house* validation (assuming that at least a few data points at the targeted thermal-hydraulic condition are available for in-house assessment).

The idea of using existing data to inform future experiments can be visualized via window-type extrapolation mapping. For this case study, an example is given in Fig. 6, where all possible (integer) pressure values are targeted. Each row on the map corresponds to a target pressure. Given a target, two sets of data are created using data from the in-house database (i.e., all tube data): (1) a floating training set ranging from $P$ to $P+3$ MPa[a] and excluding all intervals that contain the target, and (2) a fixed validation set covering the target pressure $\pm$ 0.5 MPa. The rRMSE value in each box represents the extrapolation performance of the hybrid (RF+Liu) model for each training–validation mapping[b]. The lighter the box, the smaller its corresponding validation error (i.e., rRMSE on the validation dataset).

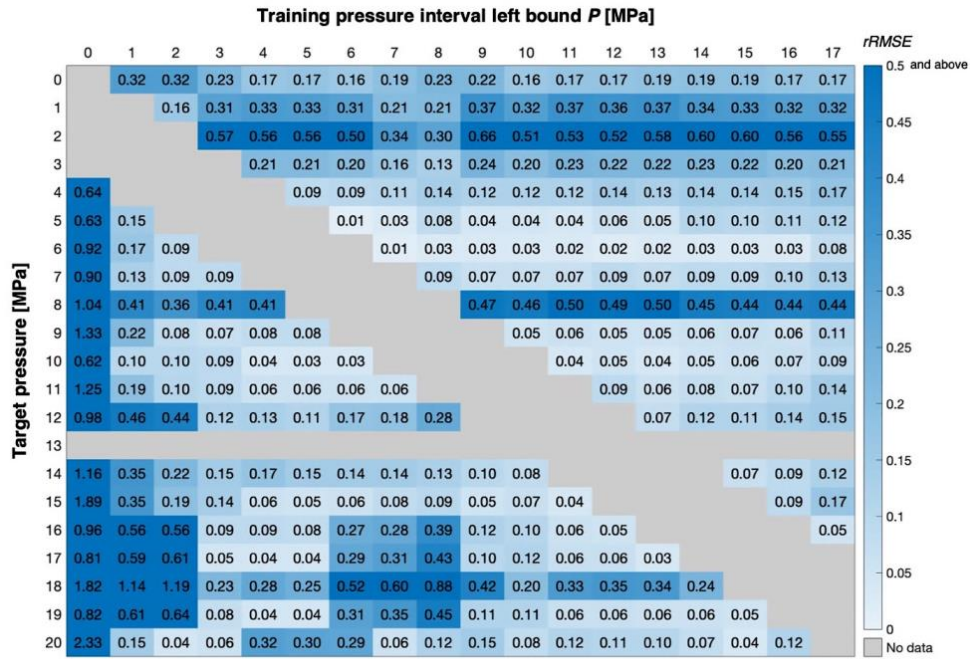

**Fig. 6.** Window-type extrapolation mapping: validation rRMSE on tube CHF with hybrid (RF+Liu) model.

---

[a] The value of $P$ varies from 0 up to 17 MPa; "$P+3$" is a case-specific choice ensuring that the training size is always larger than 50 to ensure proper learning.
[b] Note that no data is available at 13 MPa in the collected tube database.

To interpret this map, for this case study, if predictions are to be made on unseen scenarios at 15 MPa, and if a threshold rRMSE value of 0.1 is set, then new tube experiments would be deemed unnecessary as long as the existing database has covered pressure values from 4–14 MPa or from 16–19 MPa. On the other hand, with a target pressure of 8 MPa, new measurements are highly recommended, as no subset of the existing data is able to accurately predict trends and values of DNB around this pressure. One can readily tell from the map whether DK-augmented trends in a certain range of data would carry over to a different range of interest. It also indicates which thermal-hydraulic ranges should be expanded in the database to better assess the predictive capabilities of different models.

## 4 Conclusions

This paper extends the predictive capabilities of DK- and ML-based methods to improve the safety and economic competitiveness of thermal systems that depend on accurate prediction of DNB-type CHF by introducing a physics-informed, ML-aided (hybrid) framework. A comprehensive evaluation has demonstrated the superior performance of the hybrid approach as compared to standalone models, and the following conclusions can be drawn from this case study:

- By taking advantage of prior knowledge to lay the groundwork and using ML to capture the undiscovered information from the mismatch between the actual and DK-predicted target, the hybrid approach significantly outperforms the standalone models across a wide range of flow conditions with reduced scatter and unbiased parametric trends.
- One of the key features of ML tools (standalone and hybrid) is their on-the-fly extensibility of applicability domain.
- The cross-validation technique compares closely with the conventional train–test split, and it reduces error due to bias by creating more randomly generated subsets.
- Within the hybrid framework, the ML structure can be made simpler than its standalone counterpart (to save computational cost when dealing with large datasets).
- The hybrid approach results in more robust extrapolation capabilities than those obtained with standalone ML methods.
- Within the hybrid framework, while the choice of DK model seems trivial for interpolation purposes, it is more important when it comes to extrapolation (in respect of the ML-aided component), especially when the training dataset is relatively small.

- The proposed methodology of window-type extrapolation mapping leverages existing data to help inform whether or not new measurements are needed for a targeted thermal-hydraulic condition.

Both model and data complexity can be further extended with no extra burden on training/validation (and testing, if data size becomes much larger). The input feature dimension will likely increase if other channel geometries are included, such as rod bundles in a nuclear reactor. Engineered flow (e.g., nanofluids) and surface (e.g., oxidation, wickability, wettability) characteristics can also be included in the feature vector, as they all have exhibited significant impacts on CHF (at least at low pressures) [49–51]. Finally, areas of future interest also include making transient (i.e., time-series) predictions and quantifying uncertainties by integrating more advanced ML tools such as recurrent neural network, Bayesian neural network, quantile random forest, and even transfer learning into the hybrid framework.

The generic hybrid approach along with its associated methodology of window-type extrapolation mapping proposed in this work are suitable for a broad spectrum of applications in nuclear reactors and other thermal systems, where domain knowledge and experimental (or high-fidelity numerical) data are available. Such a versatile, user-friendly framework provides insight into how conventional engineering fields could benefit from the new era of artificial intelligence.

**Acknowledgments**

This research was supported by the Consortium for Advanced Simulation of Light Water Reactors (www.casl.gov), an Energy Innovation Hub (http://www.energy.gov/hubs) for modeling and simulation of nuclear reactors under U.S. Department of Energy Contract No. DE-AC05-00OR22725.

**Appendix A. Standalone ML vs. LUT: summary**

The best-estimate standalone NN and RF methods are compared against the well-used LUT on all 1,865 data points from tube (1,439), annulus (378), and one-side heated plate (48). As shown in Fig. A.1, while the ML training errors are very small (rRMSE = 7.2% with NN and 5.2% with RF on all data), more attention should be paid to validation performance since the goal of ML algorithms is to perform well on unseen data. The validation error differences between the approach to cross-validate on all data and then analyze for each geometry (shown as *from all data*) and the approach to cross-validate on each geometry separately (shown as *tube/annulus/plate data only*) are relatively small with one exception: plate with NN[a]. This result suggests that the predictions seem to be contained within a reasonable bound. Cross-validation (10-fold) with NN or RF (1) performs in a manner similar to LUT on tube data (LUT performs marginally better, but it is important to note that the majority of tube data used here was part of the data source for generating LUT), (2) yields reduced errors on annulus data, and (3) significantly outperforms the table method on plate data.

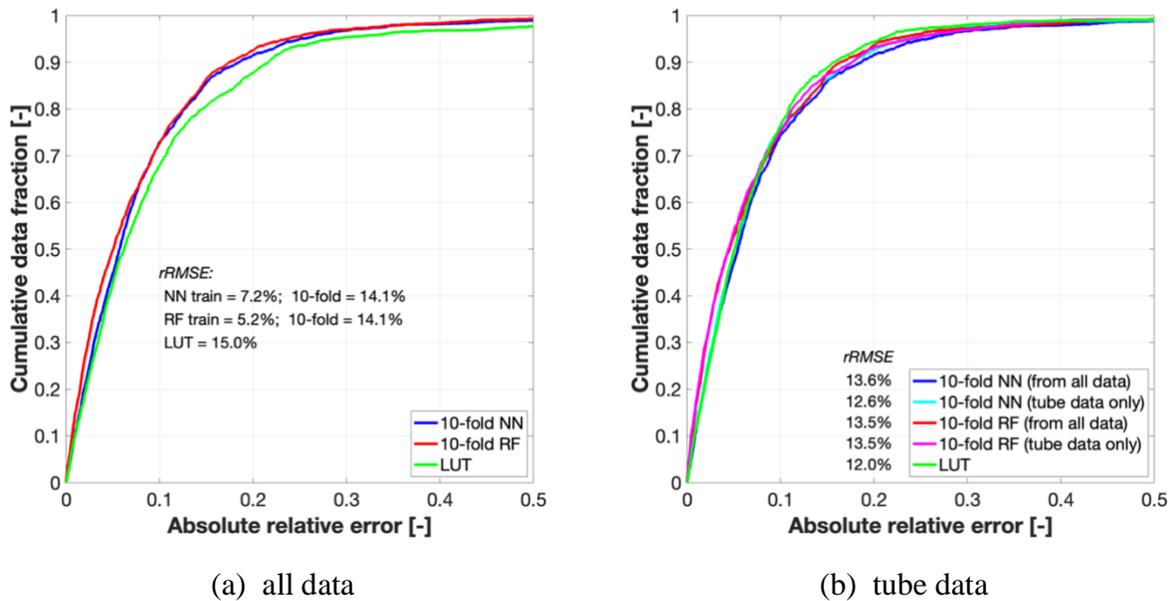

(a) all data    (b) tube data

---

[a] Given the size of the one-side heated plate dataset (only 48 collected data), *plate data only* results may not be deemed reliable for quantitative uses.

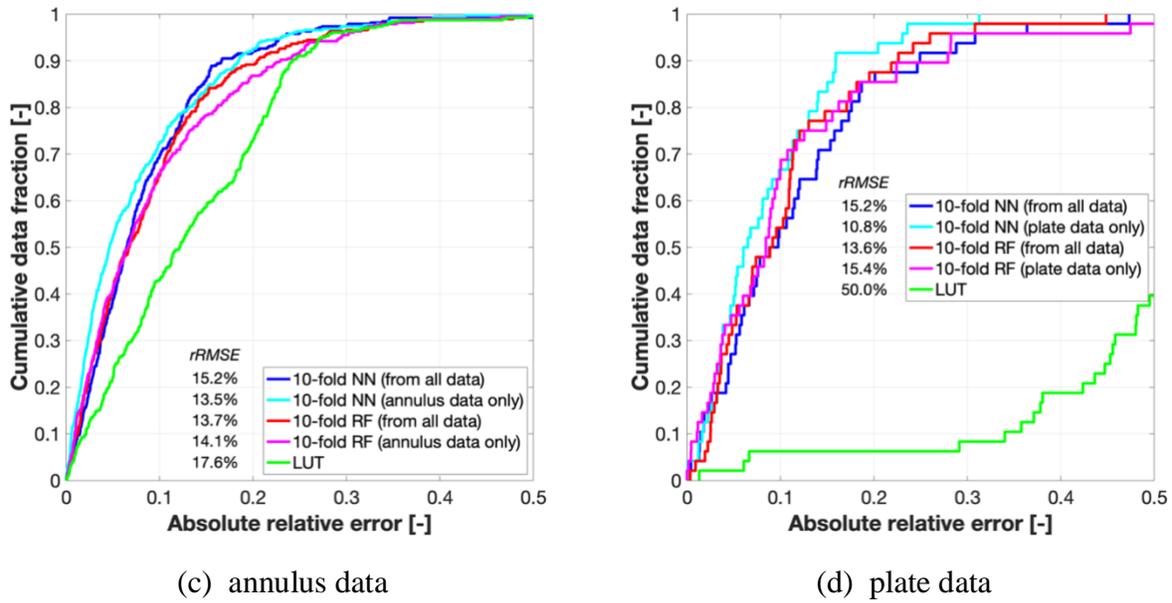

(c) annulus data          (d) plate data

**Fig. A.1.** Cumulative data fraction of absolute relative error (i.e., absolute value of relative error) with best-estimate standalone ML and LUT on
(a) all data, (b) tube data, (c) annulus data, and (d) plate data.

As can be seen in Figs. A.1d–A.2, the two ML methods agree closely with each other and with measurements of plate data: about 90% of predicted data fall within ±20% uncertainty. The LUT tends to dramatically under-predict the one-side heated plate CHF. This underestimation is consistent with previous assessments [42,43], possibly due to the much smaller heated length of the plate heater used for this work than those of the tubes used for mapping the LUT. Another explanation would be that since the LUT was originally developed for tubes only, additional correction factors may be needed for plate applications. It should also be noted that if no plate data was included during training, then similar poor performance would be expected with ML methods (i.e., standalone ML does not guarantee improved extrapolation).

In addition to the baseline case, two sets of sensitivity study have been carried out to assess the robustness and effectiveness of the standalone ML methods:

i)      *cross-validation technique*: 10-fold vs. 5-fold;
ii)     *input feature vector*: with geometry indicator (in the form of one-hot encoding, i.e., tube = [1,0,0], annulus = [0,1,0], plate = [0,0,1]) vs. without geometry indicator.

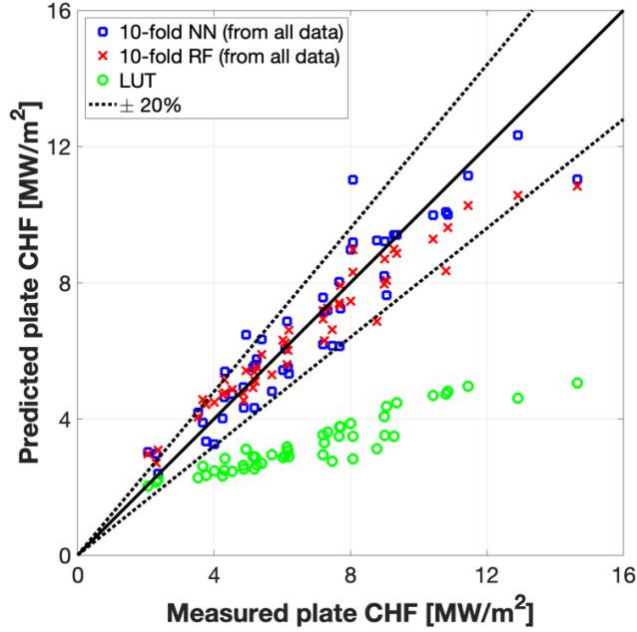

**Fig. A.2.** Predicted vs. measured plate CHF with best-estimate standalone ML and LUT.

Table A.1 summarizes the performance of three scenarios (one baseline and two sensitivity cases) on the entire dataset. The effect of the subsample number (10-fold vs. 5-fold) in cross-validation is negligible. The ML community highly recommends allocating at least 60–70% of the available data for training (i.e., $k \geq 3$ in a $k$-fold cross-validation), particularly for relatively small datasets. Adding the geometry indicator increases the input feature dimension, yet it has been shown that this feature plays a trivial role in this application. Besides, it would also require more training data to achieve better fitting and to further improve performance.

**Table A.1**

Performance of standalone ML-based CHF models on all data: sensitivity study.

| Case # | Cross-validation | Geometry indicator | NN validation rRMSE | NN validation within ±20% error | RF validation rRMSE | RF validation within ±20% error |
|---|---|---|---|---|---|---|
| *baseline* | 10-fold | no | 0.14 | 92% | 0.14 | 92% |
| *i* | 5-fold | no | 0.14 | 91% | 0.14 | 91% |
| *ii* | 10-fold | yes | 0.14 | 92% | 0.14 | 92% |

[48] T.-D. Shih, Measurement of transient critical heat flux by fluid modeling. Ph.D. thesis, Iowa State University, 1974.

[49] J. Buongiorno, L.-W. Hu, S.J. Kim, R. Hannink, B. Truong, E. Forrest, Nanofluids for Enhanced Economics and Safety of Nuclear Reactors: An Evaluation of the Potential Features, Issues, and Research Gaps, Nucl. Technol. 162 (2008) 80–91. doi:10.13182/NT08-A3934.

[50] A. Fazeli, S. Moghaddam, A New Paradigm for Understanding and Enhancing the Critical Heat Flux (CHF) Limit, Sci. Rep. 7 (2017) 5184. doi:10.1038/s41598-017-05036-2.

[51] M. Trojer, R. Azizian, J. Paras, T. McKrell, K. Atkhen, M. Bucci, J. Buongiorno, A margin missed: The effect of surface oxidation on CHF enhancement in IVR accident scenarios, Nucl. Eng. Des. 335 (2018) 140–150. doi:10.1016/J.NUCENGDES.2018.05.011.

[52] D.P. Kingma, J. Ba, Adam: a method for stochastic optimization, in: Proc. 3rd Int. Conf. Learn. Represent., San Diego, CA, USA, 2015.
30